# On the inequality of the 3V's of Big Data Architectural Paradigms: A case for heterogeneity

Todor Ivanov, Nikolaos Korfiatis, Roberto V. Zicari
(ivanov@dbis.cs.uni-frankfurt.de; korfiatis@em.uni-frankfurt.de;
zicari@informatik.uni-frankfurt.de)

*Frankfurt Big Data Laboratory*
*Chair for Databases and Information Systems*
*Institute for Informatics and Mathematics*
*Goethe University Frankfurt*
*Robert-Mayer-Str. 10, 60325, Bockenheim*
*Frankfurt am Main, Germany*

*http://www.bigdata.uni-frankfurt.de*

## ABSTRACT

The well-known 3V architectural paradigm for Big Data introduced by Laney (2011) provides a simplified framework for defining the architecture of a big data platform to be deployed in various scenarios tackling processing of massive datasets. While additional components such as Variability and Veracity have been discussed as an extension to the 3V model, the basic components (Volume, Variety, and Velocity) provide a quantitative framework while variability and veracity target a more qualitative approach. In this paper we argue why the basic 3V's are not equal due to the different requirements that need to be covered in case there exist higher demands for a particular "V". We call this paradigm heterogeneity and we provide a taxonomy of the existing tools (as of 2013) covering the Hadoop ecosystem from that perspective. This paper contributes on the understanding of the Hadoop ecosystem from both an architectural and requirements viewpoint and aims to help researchers and practitioners on the design of scalable platforms targeting different business scenarios.

**Keywords:** 3V's, Heterogeneity, Big data platforms, Big data systems architecture

# 1 Introduction

The exponential data growth in the last few years has challenged the processing and storage capabilities of modern information systems and internet platforms transforming the need to handle and manage large volumes of data to a strategic one (Chintagunta et al. 2013). This "explosion" of information has rapidly transformed the term "Big Data" (Diebold 2012) into the new hype, turning it in an essential part of the Cloud Computing Service Model (Miller 2013). To cope with this challenge and meet tight requirements such as time to process, important design improvements with respect to scalability, supporting parallel and distributed data processing have to be applied. This requires major architectural changes and the use of new software technologies like Hadoop (Apache Hadoop 2013) and its distributions from commercial vendors, whose major goal is to effectively process of massive data sets.

Theoretical definitions of what "Big Data" is and how can be utilized by organizations and enterprises has been a subject of debate (Jacobs 2009). Nonetheless, a framework that has gained considerable attention was first introduced by Laney (Laney 2001) and considers three distinctive characteristics for big data , namely: Volume, Variety and Velocity . Figure 1 (Zikopoulos and Eaton 2011) summarizes the dynamics and interconnection between these characteristics as a case of interconnected stages, known as the 3Vs of Big Data.

On the 3V model, each stage is interconnected and runs as an input to the subsequent one. The volume represents the evergrowing amount of data in petabytes, exabytes, zettabytes and yottabytes, which is generated in today's "Internet of things" and challenges the current stage of storage systems. On the other hand, the variety of data produced by the multitude of sources like sensors, smart devices and social media in raw, semi-structured, unstructured and rich media formats is further complicating the processing and storage of data. Finally, the Velocity aspect describes how quickly the data is retrieved stored and processed. This is becoming more and more of a burden for the current systems as they are not suited to deal with different not always defined formats, with increasing size, and having varying processing time requirements. From an information processing perspective, the three characteristics together describe accurately what Big Data is and the new challenges that it presents to the backend systems.

While the 3Vs model provides a simplified framework which is well understood by researchers and practitioners, our perspective is that the simplified representation of the data processes described in that, can lead to major architectural pitfalls on the design of big data platforms. A particular issue that we take into account is the cost factor or the value that derives from the utilization of the 3Vs model in the context of a business scenario. Our intuition is that since business operations are not equal in any vertical market, the influence of the 3Vs in a "Big Data" implementation process is not the same. Taking this into account we are using the 3V framework to address particular cases of different requirements and how this can be saturated on top of an existing infrastructure considering the cost factors associated with systems operations and maintenance. We refer to this architectural paradigm as *Heterogeneity* and we elaborate on the current technical solutions and architectures of the Hadoop ecosystem that can make such an architectural paradigm feasible, in particularly in large private clouds. In order to address that, we provide a classification of the tools comprising the existing Hadoop based ecosystem (as of 2013) in (a) system management, (b) platform and (c) application level. Thus the contribution of this paper is to provide a taxonomy of the Big Data Ecosystem in relation with Volume, Variety and Velocity capabilities and highlight to researchers and practitioners the heterogeneity aspect sourced by the different business domains and application scenarios.

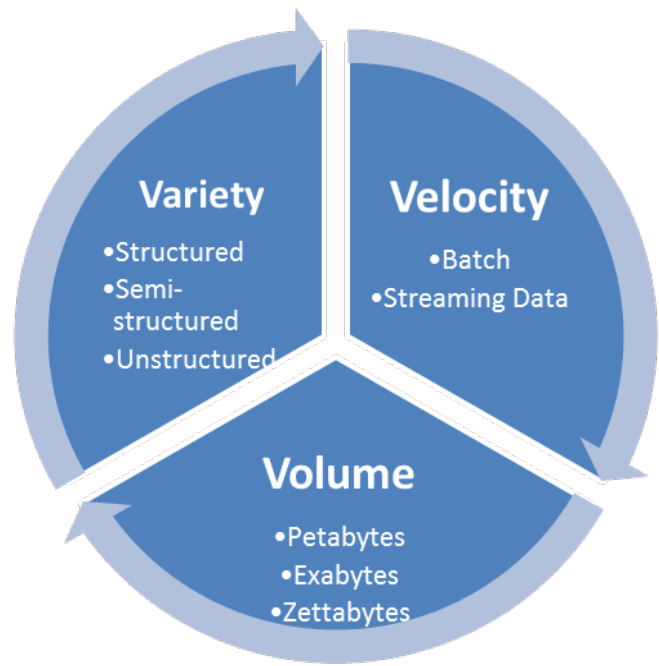

**Figure 1** IBM Big Data characteristics – 3V. Adopted from *(Zikopoulos and Eaton 2011)*

To this end this paper is structured as follows. Section (2) elaborates on challenges of big data platforms and in particular on the inequality of the 3Vs and the value dimension. Section (3) discusses heterogeneity as a feasible architectural paradigm by providing an example big data architecture. A classification is provided in three different levels considering the hardware level as invariant since Hadoop architectures are built on parallelization of commodity hardware. We provide a discussion on Section (4) with open issues on hardware system management and platform level. We conclude our perspective on Section (5) with open issues and challenges for future research.

## 2    Challenges in Big Data platforms

The Cloud Computing providers are the one trying to solve all the Big Data challenges (Zicari 2012; Zicari 2013) and to offer the customers wide variety of services through which they can reach their data insights. However, customer requirements are growing with the same pace as their data and the need to get faster insights on which to base their business decision is be-coming more crucial. This pushes the cloud vendors to optimize and improve their cloud platforms for better storage, processing and energy efficiency. With the popular "pay-per-use" service which they offer the effective resource utilization that meets the SLAs is the major factor in their cost model. Mozafari and colleagues (Mozafari et al. 2013) have identified three key needs for cloud database services to be appropriate: (i) pricing schemes that are based reflect their operational costs but are also simple and intuitive to users, (ii) performance efficient mechanisms to isolate the performance of tenants from each other, while allowing soft-sharing of resources, and (iii) workload-specific tuning for each tenant. Nevertheless, satisfying all these design criteria by implementing "One size" platform for storing and processing large data sets to address the different customer needs is a big challenge. The DBMS+ approach proposed by Lim et al. (2013) and used to build their Cyclops platform outlines these challenges: (i) interaction with its internal systems, (ii) how to select the systems to integrate, (iii) how to select the most suitable execution plan, (iv) how to provision resources,(v) how the data is stored, and (vi) what application execution requirements to support. The Cyclops system integrates a centralized streaming system (Esper 2013), a distributed streaming system (Storm 2013) and a distributed batch processing system (Apache Hadoop 2013). The architecture complexity and technical requirements that such platforms should meet are immense which makes them very expensive to be build and administered. Such platforms can be implemented only by big cloud providers like Amazon, Rackspace, GoGrid, IBM, Microsoft Azure and Google, who manage their own data centers, develop specific applications and offer them as services. This is not the case with smaller companies, public and private institution that need private cloud solutions. They also have to meet a strict regulations in terms of data privacy and security as well as specific regulations enforced by laws or other monitoring organizations. Similar to the big cloud providers their major concerns are cost, space, energy and management efficiency as well as the initial prices for software and hardware acquisitions. In that context viable architectures need to consider a "value" dimension on their implementation approach. We outline this in the section that follows.

## 2.1 Inequality of the Big Data Characteristics: The value component

In traditional storage systems volume (size) defines the amount of data which the system can manage whereas the velocity (speed) defines the speed with which this data is processed. This can be different based on different architectures. For example in OLAP systems the data size can become very large but the data processing speed is slow whereas in OLTP systems the workload size is much smaller, but the data should be processed much faster. The variety (structure) is of no concern in such systems as the data format is known in advance and most of the times is described very precisely in a pre-defined schema. However, in the case of Big Data the emerging data variety is starting to transform the system requirements and question the storage efficiency of existing systems.

A particular architectural requirement is that a system from its foundations should be capable of handling increasing data volume, high or dynamically changing velocity and high variety of data formats. The exact value of each of the 3Vs can vary depending on the industry-specific requirements and the current infrastructure platforms should be able to deal with any combination of the 3Vs. This dynamically changing inequality of the 3Vs depicts in a very abstract way the actual challenge of the Big Data. In technical terms fully-supporting the 3Vs in a system opens even more questions. Nonetheless, apart from the 3Vs, which represent the quantitative characteristic of Big Data system, there are additional qualitative characteristics like *Variability* and *Veracity*. The Variability aspect defines the different interpretations that a certain data can have when put in different contexts. It focuses on the meaning of the data instead of its variety in terms of structure or representation. The Veracity aspect defines the data accuracy or how truthful it is. If the data is corrupted, imprecise or uncertain, this has direct impact on the quality of the final results. Both variability and veracity have direct influence on the qualitative value of the processed data. The real value obtained from the data analyzes also called data insights is another qualitative measure which is not possible to define in precise and deterministic way. However, the term value is not unambiguous and can be understood as the cost value of the system. This is the case in the work of Baru and colleagues (Baru et al. 2013a; Baru et al. 2013b), where they define ***value*** as the 4th V, which in the context of benchmarking is seen as a cost-based metric(price/performance) based on the three quantitative characteristics. Actually, in this context value is not a new dimension on the framework but a function that combines the other three so that a change in each compontent will affect the value

$$value = f(volume, variety, velocity)$$

Value is actually very important metric for every Cloud and Big Data service provider (Greenberg et al. 2008). It determines how efficient is a platform in terms of price per computation unit and how effectively in terms of data storage. This metric includes the costs for hardware, maintenance, administration, electricity, cooling and space. In other words, all essential elements for building an enterprise infrastructure are included in the definition of the value (cost factor) characteristic. Clearly this characteristic is the most complex as it is defined by the 3Vs together with additional factors that they influence. Therefore, because of this complexity the Big Data platforms are difficult to benchmark and compare in terms of performance and price. Currently, the development of a Big Data benchmark is in progress and will by defined by multiple industry cases as described in (Baru et al. 2013a; Baru et al. 2013b). Currently as of 2013, there is an urgent need of standardized test workload against which all software vendors can test their software. Inspired by the latest benchmark specifications of TPC-DS(TPC 2013a) and TPC-VMS(TPC 2013b), the Big Data Top100 Project(2013a) just recently presented the BigBench benchmark (Ghazal et al. 2013). The proposed end-to-end big data benchmark represents a data model that simulates all the 3Vs characteristics of a big data system together with a synthetic data generator for structured, semi-structured and unstructured data. The structured part of the retail data model is adopted from the TPC-DS benchmark and further extended with semi-structured (registered and guest user clicks) and unstructured (product reviews). The BigBench raw data volumes can be dynamically changed based on a scale factor. The simulated workload is based on set of 30 queries covering the different aspects of big data analytics proposed by McKinsey (Manyika et al. 2011).

## 2.2 System Variety and Complexity

Monash(Curt 2013) addresses the problem that there is no single data store that can be efficient for all usage patterns, something that Stonebraker discussed in detail (Stonebraker et al. 2007) and proposed his own taxonomy of database technologies. Interestingly enough, one of the platforms has Google style architecture and looks very similar to Apache Hadoop. However, the message here is that the illusion of having one general purpose system that can handle all types of workloads is not realistic and especially today with the dynamic change of the Big Data characteristics. The fact that in the recent years emerged many new kinds of storage systems like the NoSQL and NewSQL initiatives, MapReduce-based systems, Hybrid OLAP-OLTP systems, in-memory systems and column-based only prove that it is a challenge to have a single system. Most of the approaches in these new systems are inspired by the inefficiency and complexity of the current storage systems. In addition to that, the advancements in hardware and particularly the increase of main memory sizes, the rapid embracement of the Solid State Disks and the multicourse processors which together brought the prices of enterprise hardware down to the level of commodity machines.

In his paper, Cattell (Cattell 2011) identifies six key features of the NoSQL data stores namely: (1) the ability to horizontally scale "simple operation" throughput over many servers; (2) the ability to replicate and to distribute (partition) data over many servers; (3) a simple call level interface or protocol (in contrast to a SQL binding); (4) a weaker concurrency model than the ACID transactions of most relational (SQL) database systems; (5) efficient use of distributed indexes and RAM for data storage; and (6) the ability to dynamically add new attributes to data records. Similarly, Strauch (Strauch et al. 2011) summarizes all the motivations behind the emergence of the NoSQL databastores among which are the avoidance of unneeded complexity and expensive object-relational mapping, higher data throughput, ease of horizontal scalability (do not rely on the hardware availability) and offer new functionalities more suitable for cloud environments in comparison to the relational databases. Additionally, he presents extensive classification and comparison of the NoSQL databases by looking into their internal architectural differences and functional capabilities.

Industry perspectives such as the one's advocated by Fan (Charles 2012) view emerging systems as a transformation between the traditional relational database systems working with *CRUD* (Create, Read, Update, Delete) data and the *CRAP* (Create, Replicate, Append, Process) data. His major argument is that the CRUD (structured) data is very different from the CRAP (unstructured) data because of the new Big Data characteristics. The new semi-structured and unstructured data is stored and processed in near real-time and not really updated. The incoming date streams are appended. Therefore, CRAP data has very different characteristics and is not appropriate to be stored in the relational database systems.

Marz(Nathan 2012) discusses the problem of mutability of the existing database architectures. More specifically the Update and Delete in the CRUD data which actually change the data consistency and lead to undesired data corruption and data loss caused by human interaction. Actually he suggests a new Big Data architecture called *Lambda Architecture* (Marz and Warren 2012) which major principles are human fault-tolerance, data immutability and recomputation. By removing the U and D from CRUD and adding the append functionality similar to the Charles Fan, the data immutability is assured. The raw data is aggregated as it comes and sorted by timestamp which greatly restricts the possibility of errors and data loss caused by human fault-tolerance. The re-computation or data processing is done simply by applying a function over the raw data (query). In addition to that the architecture supports both batch and real-time data processing.

Inspired by Google's MapReduce paper (Dean and Ghemawat 2008), Hadoop based systems have been growing in adoption due to their performance, scalability and fault-tolerant features, as well as their distributed parallel processing abilities. The fact that such systems can be built on commodity hardware and its licensing model provide an important advantage over commercial vendors. In a similar spirit of innovation, most of the new infrastructure architectures try to solve only a predefined set of problems, bound to specific use case scenarios and ignore the other general system requirements. Therefore, a typical design approach is to combine two or more system features and build a new hybrid architecture which improves the performance for the targeted use case, but adds an additional complexity. HadoopDB (Abouzeid et al. 2009) is such hybrid system, trying to combine the best features of the MapReduce-based systems and the traditional analytical DBMS, by integrating PostgreSQL as the database layer, Hadoop as the distributed communication layer and Hive as a translation layer. Other systems just iteratively improve an existing platform like Haloop (Bu et al. 2010; Bu et al. 2012) and Hadoop++ (Dittrich et al. 2010) which further improve the Hadoops' scheduling and caching mechanisms as well as indexing and joining processing. Also Starfish (Herodotou et al. 2011) extends Hadoop by enabling it to automatically adapt and self-tune depending on the user workload and in this way provide better performance. A comprehensive survey by Sakr et al. (2013) on the family of MapReduce frameworks provides an overview of approaches and mechanisms for large scale data processing.

In a recent work Qin and colleagues (Qin et al. 2013) identify the MapReduce computing model as a de-facto standard which addresses the challenges stemming from the 3V Big Data characteristics. Furthermore, the authors divide the enterprise big data platforms in three categories: (1) Co-Exist solutions; (2) SQL with MapReduce Support solutions; and (3) MapReduce with SQL Support solutions. In the first category they put IBM Big Data Platform and Oracle Big Plan as both offer end-to-end solutions consisting of several data management and processing components. In the second category fall systems integrating Hadoop support like PolyBase (DeWitt et al. 2013), EMC Greenplum and TeraData Aster Data. In the last category fall Hadoop systems that integrate SQL support using Drill, Hive, Hortonworks Stinger, Cloudera Impala and similar.

### 3 Heterogeneity as a feasible architectural paradigm

The concept of system heterogeneity has been of interest for both researchers and industries for many years and still remains a hot topic. There are number of motivations behind the use of heterogeneous platforms, but recently a few of them have become very essential: i) many new hardware capabilities – multi-core CPUs, growing size of main memory and storage and different memory and processing accelerator boards such as GPUs, FPGAs and caches; ii) growing variety of data-intensive workloads sharing the same host platform; iii) complexity of data structures; iv) geographically distributed server locations and v) higher requirements in terms of cost, processing and energy efficiency as well as computational speed.

### 3.2 Related Work

There are multiple studies and surveys on heterogeneous systems which try to classify and order them according to their properties and workloads for which they are best suitable. However, because of the rapidly changing architecture both in terms of hardware and software the concepts and levels of heterogeneity in the platforms evolve with the time.

For example, a survey by Khokhar et al. (1993, pp: 19) defines Heterogeneous Computing (HC) as "*..A well-orchestrated, coordinated effective use of a suite of diverse high-performance machines(including parallel machines) to provide fast processing for computationally demanding tasks that have diverse computing needs..*". In addition, the authors discuss multiple issues and problems stemming from system heterogeneity among which are three very general: i) "the types of machines available and their inherent computing characteristics"; ii) "alternate solutions to various sub-problems of the applications" and iii) "the cost of performing the communication over the network".

In another survey on Heterogeneous Computing by Ekmecic et al. (1996) the authors discuss the heterogeneous workloads as a major factor behind the need of heterogeneous platforms and divide the heterogeneous computing in three essential phases: i) parallelism detection, ii) parallelism characterization and iii) resource allocation. The Parallelism detection phase is responsible for discovering parallelism inside every task in a heterogeneous application. The Parallelism characterization estimates the computation parameters like most suitable execution mode and time as well as the amount and cost for communication for each task of an application. The Resource allocation determines an exact place and moment of execution for every task taking into account certain performance metrics and other constraints like overall machine cost for the execution. Basically, the phases describe in a more abstract way today's concept of cloud computing.

In a similar study Venugopal et al. (2006), present a taxonomy of Data Grids, and highlight heterogeneity as an essential characteristic of data grid environments and applications. Furthermore, they briefly mentioned that heterogeneity can be split in multiple levels like hardware, system, protocol and representation heterogeneity, which resemble very much the presented Data Grid layered architecture.

Interestingly enough, the characteristics of today's Big Data platforms as well as the challenges and problems that they represent are very similar to the one discussed in Heterogeneous Computing and Data Grid environments. Therefore, it is a logical step to look in more detail at the concept of system heterogeneity and investigate how it is coupled with the Big Data characteristics.

Lee and colleagues (Lee et al. 2011) discuss the importance of heterogeneity in cloud environments and suggest a new architecture and techniques to improve the performance and cost-effectiveness. They propose an architecture consisting of 1) long-living core nodes to host both data and computation and 2) accelerator nodes that are added to the cluster temporarily when additional power is needed. Then the resource allocation strategy dynamically adjusts the size of each pool of nodes to reduce the cost and improve utilization. Additionally, they present a scheduling scheme, based on the job progress as a shared metric, which provides resource fairness and improved performance.

In another study Mars et al. (2011) investigated micro-architectural heterogeneity in warehouse-scale computer (WSC) platforms, In that the authors present a new metric called opportunity factor that approximates the application's potential performance improvement opportunity relative to all other applications and given the particular mix of applications and machine types on which is running. Next, they introduce opportunistic mapping which solves the optimization problem of finding the optimal resource mapping for heterogeneity-sensitive applications. Using this technique the performance of real production cluster improved with 15%, but can go up to 70%.

The number of studies related to heterogeneity is growing along the conceptual relevance and challenging problems that it brings. However, most of the research is done for a particular heterogeneity level and does not take into account the overall system architecture. Therefore, in the next sections we try to give a more generic view and demonstrate the multiple points where the concept of heterogeneity appears in the context of a Big Data platform.

### 3.3 An example heterogeneous Big Data Architecture

In order to elaborate further on the heterogeneity of the different system layers we introduce an example architecture in order to discuss how different business scenarios and workloads can be addressed. Figure 2 depicts an in-depth overview of such Big Data platform suitable for Analytical, Business Intelligence, ETL (Extract-transform-load) and Reporting workloads, aggregating large data sizes and processing them in a batch or near-real time manner.

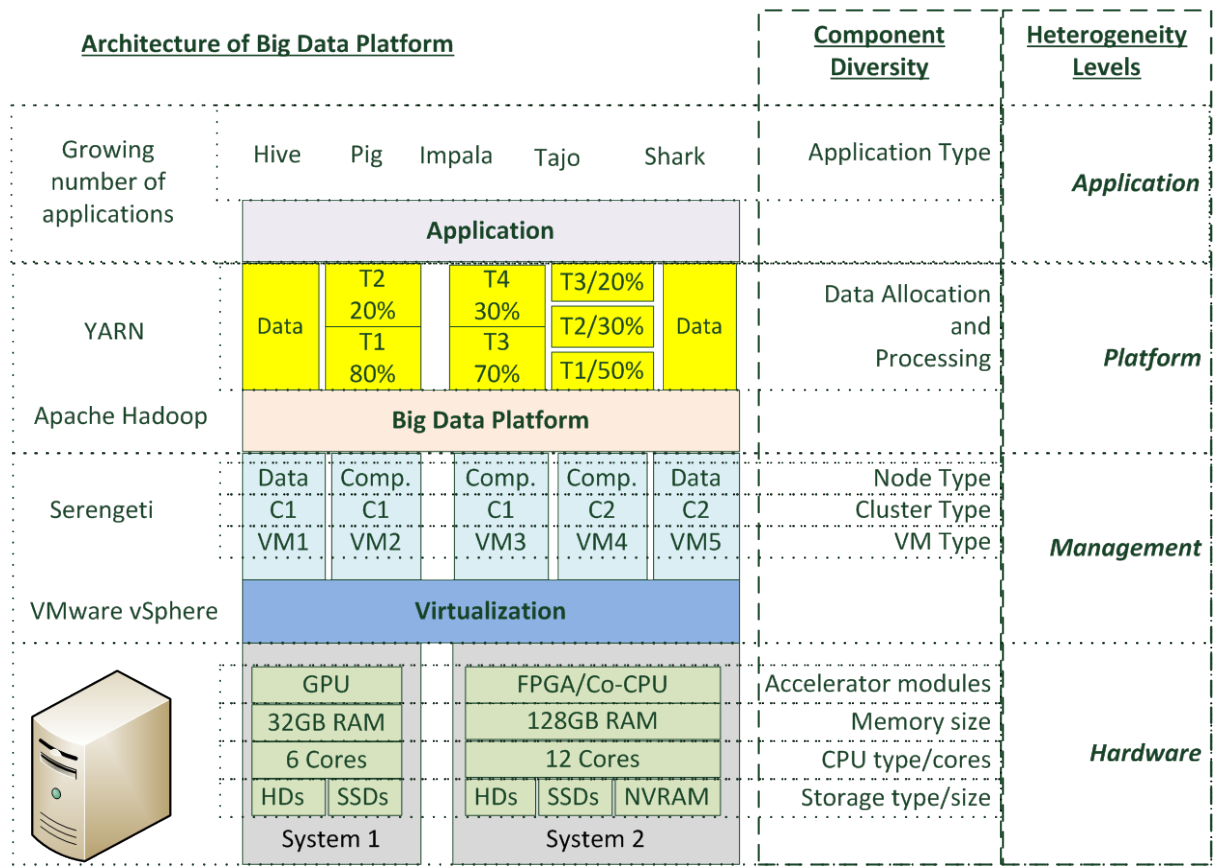

**Figure 2** Exemplary Architecture of a Big Data Platform

Bottom up, the first platform layer is the hardware level represented in our scenario by Systems 1 and 2. Both servers consist of very different hardware components. As depicted, the systems differ in terms of HDs and SSDs sizes with additional NVRAM, CPU type and number of cores, main memory size as well as additional processing units like GPU for system 1 and FPGA board and co-processor for System 2. The basic idea behind is to show that the server architectures can greatly vary by offering different processing capabilities, which can improve the performance of particular workload. Next is the virtualization layer, consisting of the VMware vSphere(VMware 2013a) cloud platform, which is responsible for the resource allocation and management of the underlying hardware. It distributes the hardware resources between the virtual machines, which represent the nodes in a virtualized cluster. The management of the Hadoop cluster is further improved by the use of the Serengeti(VMware 2013b) server, which allows starting, stopping and pre-configuring Hadoop clusters on the fly. It is an open source project started by VMware and now integrated in vSphere as Big Data Extension, which has the goal to ease the management of virtualized Hadoop clusters. By implementing of hooks to all major Hadoop modules it is possible to know the exact cluster topology and make it aware of the hypervisor layer. This open source module is called Hadoop Virtual Extension (HVE) (VMware 2012) . What is really interesting is the new ability to define the nodes (virtual machines) as either only computing or data nodes. The above implies that some nodes are storing the data in HDFS while others are responsible for the computation of MapReduce jobs. Another very similar project called Savanna(OpenStack 2013) was recently started, as a part of the OpenStack platform, and has the goal to improve the Hadoop cluster deployment and management in cloud environments.

As depicted on the Figure 2, the virtual machines can represent cluster nodes of different type (data or computation) and be part of different logical clusters. Clusters consist of data and computation logic that is specified by the application workload. Therefore, multiple clusters can share and co-exist on the same hardware infrastructure. The clusters can share the same dataset, but still the applications running in the clusters can have completely different characteristics. The system architecture depicted on Figure 2 is just an example and the components and versions of the different applications on the layers can greatly vary. The following section presents a classification of the heterogeneity levels of a Big Data platform. Furthermore, a list of available software components together with their ability to handle the three Big Data characteristics (Volume, Variety and Velocity) is briefly described in a tabular format.

### 3.4 Classifying levels of Heterogeneity

Having provided an exemplary Big Data architecture, we proceed on discussing the different levels of heterogeneity on Hardware, System management, Platform and Application Level. For each particular level we provide an evaluation based on Volume, Variety and Velocity, highlighting the positive or negative attribute it has or naming it invariant in the case is not affected at all. We provide our taxonomy on the sections that follow.

### 3.4.1  Hardware Level

Undoubtedly recent advances in the processing and storing capabilities of the current commodity (off-the-shelf) servers have drastically improved while at the same time becoming cheaper. This reduces the overall cost of the Big Data cluster platforms consisting of thousands of machines and enables the vendors to cope with the exponentially growing data volumes as well as the velocity with which the data should be processed. However, there have been other components like FPGAs, GPUs, accelerator modules and co-processors which have become part of the enterprise-ready servers. They offer numerous new capabilities which can further boost the overall system performance such as: i) optimal processing of calculation intensive application; ii) offloading part or entire CPU computations to them; iii) faster and energy efficient parallel processing capabilities; and iv) improved price to processing ration compared to standard CPUs. Recently, there have been multiple studies investigating how these emerging components can be successfully integrated in the Big Data platforms. In (Shan et al. 2010), the authors present a MapReduce framework (FPMR) implemented on FPGA that achieves  31.8x speedup compared to CPU-based software system.

Diversifying the core platform components motivates the investigation of the concept of heterogeneity on a hardware level and the new challenges that it introduces. Using the right hardware modules for each application will be crucial for reaching the optimal price/performance ratio.

### 3.4.2 System Management Level

The system layer is positioned directly above the hardware level and is responsible for the management and optimal allocation and usage of the underlying hardware components. There are multiple ways to achieve this: i) directly installing operating system; ii) using virtualization technology; and iii) hybrid solution between OS and virtualization. In the recent years, virtualization has become the standard technology for infrastructure management both for bigger cloud and datacenter providers as well as for smaller private companies. However, along with the multiple benefits that virtualization brings, there are also new challenges. The co-location of virtual machines hosting different application workloads on the same server makes the effective and fair resource allocation problematic. Also the logical division of virtual machines with similar characteristics is not always possible. In the case of Big Data platforms with changing workloads, it is difficult to meet the network and storage I/O guarantees. These are just few examples, which illustrate the complexity of the virtualized environment and motivate the existence of heterogeneity on the management level. An extensive list of components defining the management level of heterogeneity is provided in Table 1.

**Table 1** Management Heterogeneity

| Component | Description | Volume | Variety | Velocity |
|---|---|---|---|---|
| **Serengeti** (VMware, 2012) | *It is an open-source project, initiated by VMware, to enable the rapid deployment of Hadoop (HDFS, MapReduce, Pig, Hive, and HBase) on a virtual platform (vSphere).* | Invariant | Invariant | Invariant |
| **Savanna** (OpenStack 2013) | *It aims to provide users with simple means to provision a Hadoop cluster at OpenStack by specifying several parameters like Hadoop version, cluster topology, nodes hardware details and a few more.* | Invariant | Invariant | Invariant |
| **Mesos** (Hindman et al. 2011) | *A cluster manager that provides efficient resource isolation and sharing across distributed applications, or frameworks like* | Invariant | Invariant | Invariant |
| | *Hadoop, MPI, Hypertable, Spark, and other applications.* | | | |
| **Ambari** | *It provides an intuitive, easy-to-use Hadoop management web UI backed by its RESTful APIs for provisioning, managing, and monitoring Apache Hadoop clusters* | Invariant | Invariant | Invariant |
| **Whirr** | *Whirr is a set of libraries for running cloud services. It provides a cloud-neutral way to run services, a common service API and can be used as a command line tool for deploying clusters.* | Invariant | Invariant | Invariant |
| **ZooKeeper** (Junqueira and Reed 2009; Hunt et al. 2010) | *A centralized service that enables highly reliable distributed coordination by maintaining configuration information, naming, providing distributed synchronization, and group services.* | Achieves high throughput for coordination services | Invariant | Achieves low latency and performance |

Additionally, projects like Serengeti and Savanna, which major goal is to automate the control and resource management of virtualized Hadoop distributions, introduce new functionalities for cluster management. It is possible to define your virtual machines as data or computation nodes, something which is not possible in the purely physical environment where both Task-Tracker and DataNode run on the same machine. Furthermore, you can have many logical clusters on the same hardware platform, representing completely different application workloads. Similar multi-cluster architecture with core nodes (running both TaskTracker and DataNode) and transient nodes (running only TaskTracker) is proposed by (Ghit et al. 2012). The authors investigate different policies for effective resource management in MapReduce multi-cluster systems and identify four major types: performance, data, failure and version isolation, which can be divided in two groups: inter-cluster isolation (across different physical clusters) and intra-cluster isolation (within single physical cluster).

### 3.4.3 Platform Level

The platform layer represents the actual Big Data application environment which is responsible for the provision of general data and processing capabilities. In the example above, this is the Apache Hadoop framework consisting of the HDFS and YARN (MapReduce 2.0) core components. HDFS is responsible for the data storage whereas YARN is for the processing and resource allocation between the jobs. However, it is possible to run in parallel both YARN and MapReduce jobs for different applications. Recently Yahoo released Storm-YARN (Yahoo 2013) application which combines the advantages of both applications: real-time (low-latency) and batch processing. It enables Storm applications to utilize the Hadoop resources managed by YARN, which will offer new abilities for faster and more optimal data processing. The Spark platform introduced by Zaharia and colleagues (Zaharia et al., 2012; Zaharia et al., 2010) is built on top of HDFs and introduces the concept of Resilient Distributed Datasets (RDDs). RDDs are fault-tolerant, parallel data structures that let users explicitly persist intermediate results in memory, control their partitioning to optimize data placement, and manipulate them using a rich set of MapReduce-like parallel operations (iterative machine learning algorithms and interactive data analytics). Table 2 provides a comparison of the heterogeneity presented on the management tools rapidly gaining adoption.

**Table 2** Platform Heterogeneity

| Component | Description | Volume | Variety | Velocity |
|---|---|---|---|---|
| **HDFS** (Hadoop Distributed File System) (Borthakur 2008) | *A distributed file system that provides high-throughput access to application data.* | The essential part of the architecture that enables the management of exponentially growing volumes of data. | Invariant | Can cope with the speed up to a certain point. Use of additional tools is further beneficial. |
| **MapReduce** (Dean and Ghemawat 2008) | *A YARN-based system for parallel processing of large data sets.* | Very suitable for large amounts of data. | Invariant | Batch processing is not suitable for near real-time processing. |
| **YARN** (Yet Another Resource Negotiator)(Vavilapalli 2013) | *A framework for job scheduling and cluster resource management.* | As successor of MapReduce, it is specifically designed to work with large sets of data. | Invariant | Achieves better speed and guarantees for processing time in comparison to MapReduce. |

| | | | |
|---|---|---|---|
| **Storm**<br><br>(Leibiusky et al. 2012) | *An open source distributed real-time computation system. Storm makes it easy to reliably process unbounded streams of data, doing for real-time processing what Hadoop did for batch processing.* | Accepts large streams of data. | Invariant | Designed to process data with high velocity in real-time. |
| **Storm-YARN**<br><br>(2013b) | *It enables Storm applications to utilize the computational resources in a Hadoop-YARN cluster along with accessing Hadoop storage resources such as HBase and HDFS.* | Optimal solution for large data sets and streams. | Invariant | Combines batch processing and real-time processing. |
| **Spark**<br><br>(Zaharia et al., 2012; Zaharia et al., 2010) | *An open source cluster computing system that aims to run programs faster by providing primitives for in-memory cluster computing. Jobs can load data into memory and query it repeatedly much more quickly than with disk-based systems like Hadoop MapReduce.* | Targets the processing of large data sets | Invariant | Very suitable for interactive data analytics |
| **Oozie**<br><br>(Islam et al. 2012) | *A workflow scheduler system to manage Apache Hadoop jobs.* | Number of jobs can influence the performance | Invariant | Batch oriented processing in form of workflows of actions |
| **Chukwa**<br><br>(Boulon et al. 2008; Rabkin and Katz 2010) | *A data collection system for managing large distributed systems.* | Collects large data sets | Designed for semi-structured data like log files | Relies on MapReduce (stream processing) |
| **Tez** | *A general-purpose resource management* | Designed for large-scale queries and | Invariant | Designed to speeds up data |
| (2013c) | *framework which allows for a complex processing of directed-acyclic-graph of tasks and is built atop Hadoop YARN.* | data sets | | processing across both small-scale, low-latency and large-scale, high-throughput workloads. |
| **REEF** ( Retainable Evaluator Execution Framework) (Chun et al. 2013) | *REEF framework builds on top of YARN to provide crucial features (Retainability, Composability, Cost modeling, Fault handling and Elasticity) to a range of different applications.* | REEF is designed for large data sets, but is not dependent on data model and semantics of the system. It optimizes the communication and data movement. | Invariant | Designed to improve scalability, fault-tolerance and compensability of jobs which results in better performance. |

All these platforms emerge due to the different types of workloads that require new system capabilities. The platform should offer the computational and storage functionalities to the upper application layer. Therefore, the importance of understanding heterogeneity on this platform level is very essential for the successful management and processing of large datasets.

### 3.4.4 Application Level

Application level of heterogeneity acts as a logical extension of the major requirements of data platforms. Satisfying all the Big Data characteristics requires the platform to support all types of components starting from the data retrieval, aggregation and processing including data mining and analytics. Therefore, applications with very different characteristics should be able to run effectively co-located on the same platform, which should further guarantee optimal resource and functionality management, fair scheduling and workload isolation. To achieve these, the variability in their requirements has to be taken into account and all the platform layers should be synchronized accordingly. The underlying system layers should provide these services so that the application layer can successfully host the variety of applications. In our case, we are interested in data-intensive applications like Hive, which provide warehouse functionality on top of MapReduce-style systems. Table 3 represents a more extended list of such data-intensive applications which are able to process both structured and unstructured data.

**Table 3** Application Heterogeneity

| Component | Description | Volume | Variety | Velocity |
|---|---|---|---|---|
| **Hive** (Thusoo et al. 2009; Thusoo et al. 2010) | *A data warehouse infrastructure that provides data summarization and ad hoc querying.* | Very good for sequential processing of large data sets. | Invariant | Not suitable for high velocity processing. |
| **Pig** (Olston et al. 2008; Gates et al. 2009) | *A high-level data-flow language and execution framework for parallel computation.* | Very good for sequential processing of large data sets. | Invariant | Not suitable for high velocity processing. |
| **Impala** (Kornacker and Erickson 2012) | *It is an open source Massively Parallel Processing (MPP) query engine that runs natively on Hadoop, enabling users to issue low-latency SQL queries to data stored in HDFS and HBase without requiring data movement or transformation.* | Implemented its own processing infrastructure very similar to MapReduce. | Invariant | Improves the Hive processing to support near real-time. |
| **Tajo** (Choi et al. 2013) | *A relational and distributed data warehouse system for Hadoop,that is designed for low-latency and scalable ad-hoc queries, online aggregation and ETL on large-data sets by leveraging advanced database techniques.* | Targeting ETL and various big data set transformations | Invariant | Designed for low-latency by providing local query engines and optimized query planning compared to Hive |
| **Shark** (Engle et al. | *A fully Hive-compatible data* | Designed for massive data | Invariant | Designed for deep analysis |
| 2012; Xin et al. 2012) | *warehousing on top of Spark system that can run 100x faster than Hive.* | sets, much more scalable and flexible than Data warehouses | | and interactive, ad-hoc, and exploratory queries |
| **Mahout** (Owen et al. 2011) | *A scalable machine learning and data mining library.* | Designed to work with big datasets by sitting on top of MapReduce. | Can work with all types of data after small adjustments for preprocessing. | Relying on MapReduce jobs speed |
| **Drill** (Hausenblas and Nadeau 2013) | *An open-source software framework (inspired by Google's Dremel) that supports data-intensive distributed applications for interactive analysis of large-scale datasets.* | Designed for large-scale datasets | Invariant | Supports low-latency interactive analysis |
| **HCatalog** (now part of Hive) (Capriolo et al. 2012) | *A set of interfaces that open up access to Hive's metastore for tools inside and outside of the Hadoop grid.* | Invariant | Invariant | Invariant |
| **Sqoop** (Ting and Cecho 2013) | *A tool designed for efficiently transferring bulk data between Apache Hadoop and structured datastores such as relational databases.* | Suitable for large bulk-data sets | Designed to extract and import structured data | Invariant |
| **Tika** (Mattmann and Zitting 2011) | *A toolkit that detects and extracts metadata and structured text content* | Invariant | Suitable for structured text data | Invariant |

| | | | |
|---|---|---|---|
| | *from various documents using existing parser libraries.* | | | |
| **Flume** | *A distributed, reliable, and available service for efficiently collecting, aggregating, and moving large amounts of log data.* | Designed for large streaming data flows | Invariant | Enables fast data retrieval. Very suitable for stream and event based processing. |
| **Avro** | *A data serialization system. Avro provides: 1) rich data structures; 2) a compact, fast, binary data format; 3) a container file, to store persistent data; 4) remote procedure call (RPC) and 5) simple integration with dynamic languages.* | Suitable large data files | Relies on schema which defines the underlying data format | Invariant |
| **HBase** (George 2011) | *A scalable, distributed database that supports structured data storage for large tables.* | Designed for large data sets | Suitable for structured and semi-structured data | Provides random, real-time read/write access |

## 4 Discussion

The major factors defining platform architecture are in the relations between the quantitative Big Data characteristics – volume, variety and velocity. Implementing these relations successfully in terms of technical requirements directly impacts the complexity and energy consumption of the end system. An additional consequence of the quantitative characteristics in the system heterogeneity is incorporated on multiple layers which provides on the(i) the ability to tune it and run diversity of workloads, (ii) reduces the system costs in terms of hardware and energy consumption and (iii) offers scalable platform that can run new types of workloads.

However this can make the system complex and difficult to configure and manage as well as prone to non-trivial bugs.

In terms of technical challenges there are many questions in both practical and scientific areas that are still not clear or yet to be answered. Table 4 summarizes a set of open questions that have to be answered when building a Big Data platform for specific workload scenarios.

**Table 4** Challenging Questions

| Heterogeneity level | Questions |
|---|---|
| **Hardware** | In which case is better to use HDs, SSDs or other type of cache accelerators? Which is the best ratio (HDs/SSDs) for my workload? |
| | For which scenarios makes sense to use GPUs, Co-processors or FPGA boards in terms of cost and performance? |
| **Management** | Which Hadoop workloads are suitable for virtualized hardware? |
| | What is the optimal resource allocation in terms of vRAM and vCPU for virtualized data and computation nodes? What are the trade-offs and the rules of thumb in such cases? |
| | Which is better RAID and JBOD and for which types of workloads? What is the optimal configuration in virtualized environment? |
| | Is there a central place from where to administer, monitor and manage the entire platform? If there is no, how this can be achieved? |
| | Is there a way to save and load on demand pre-defined hardware and software configurations for the entire platform? How far is possible to automate the process of deploying of such configuration? |
| Platform | Concerning the HDFS configuration: What is the optimal chunk size and replication factor for the different types of workloads? |

In addition platform security is another major concern in relation with scalability and maintenance. Supporting and ensuring both the user and internal system security are very important for most industry sectors dealing with large data pools. Implanting the necessary secure methods into the current platforms is an essential step towards the approval and adoption of any enterprise Big Data platform. On the other hand, convincing the customer that these security measures are in place and properly working can turn to be another challenge.

## 5 Conclusions and challenges

In this paper we introduced the concept of heterogeneity in relation with the implementation of Big Data platforms and discussed how the existing tools comprising the Hadoop ecosystem adapt on the 3V's. While data Variability and Veracity are also discussed as additional dimensions on this initial model (Foster 2012; Gattiker et al. 2013), we believe that the core V's represent the basics for a more complete and systematic Big Data framework. The emergence of new analytical applications open new Big Data challenges (Zicari 2013).

These challenges are not only in relation with Data Characteristics (quality, availability, discovery and comprehensiveness), but also in terms of Data Processing (cleansing, capturing, and modeling) and Data Management (privacy, security and governance). Such an evaluation framework should be able to give the technology guidelines on how to build the best price/performance Big Data platform for a particular workload. In addition, such a platform should also be able to be evaluated with benchmark suites in order to validate that it meets the specific application requirements.